\documentclass[aps,prb,preprint,superscriptaddress,amsmath,amssymb,endfloats]{revtex4}

\usepackage{txfonts}
\usepackage{graphicx}
\usepackage{dcolumn}
\usepackage{bm}
\usepackage{color}
\usepackage[T1]{fontenc}

\newcommand{\Tc}{\ensuremath{T_{\rm c}}}
\newcommand{\GMT}{Ge$_{1-x}$Mn$_x$Te}

\begin{document}

\title{Heat-Treatment-Induced Switching of Magnetic States in the Doped Polar Semiconductor \GMT}

\author{M.~Kriener}
\email[corresponding author: ]{markus.kriener@riken.jp}
\author{T.~Nakajima}
\author{Y.~Kaneko}
\author{A.~Kikkawa}
\author{X.~Z.~Yu}
\affiliation{RIKEN Center for Emergent Matter Science (CEMS), Wako 351-0198, Japan}
\author{N.~Endo}
\affiliation{EM Business Unit, JEOL Ltd., Akishima,196-8558, Japan}
\author{K.~Kato}
\affiliation{RIKEN SPring-8 Center, Hyogo 679-5148, Japan}
\author{M.~Takata}
\affiliation{RIKEN SPring-8 Center, Hyogo 679-5148, Japan}
\author{T.~Arima}
\affiliation{RIKEN Center for Emergent Matter Science (CEMS), Wako 351-0198, Japan}
\affiliation{Department of Advanced Materials Science, University of Tokyo, Kashiwa 277-8561, Japan}
\author{Y.~Tokura}
\affiliation{RIKEN Center for Emergent Matter Science (CEMS), Wako 351-0198, Japan}
\affiliation{Department of Applied Physics and Quantum-Phase Electronics Center (QPEC), University of Tokyo, Tokyo 113-8656, Japan}
\author{Y.~Taguchi}
\affiliation{RIKEN Center for Emergent Matter Science (CEMS), Wako 351-0198, Japan}

\date{\today}

%abstract
\begin{abstract}
Cross-control of a material property -- manipulation of a physical quantity (e.g., magnetisation) by a nonconjugate field (e.g., electrical field) -- is a challenge in fundamental science and also important for technological device applications. It has been demonstrated that magnetic properties can be controlled by electrical and optical stimuli in various magnets. Here we find that heat-treatment allows the control over two competing magnetic phases in the Mn-doped polar semiconductor GeTe. The onset temperatures \Tc\ of ferromagnetism vary at low Mn concentrations by a factor of five to six with a maximum $\Tc\approx 180$~K, depending on the selected phase. Analyses in terms of synchrotron x-ray diffraction and energy dispersive x-ray spectroscopy indicate a possible segregation of the Mn ions, which is responsible for the high-\Tc\ phase. More importantly, we demonstrate that the two states can be switched back and forth repeatedly from either phase by changing the heat-treatment of a sample, thereby confirming magnetic phase-change-memory functionality.
\end{abstract}

\maketitle

%Introduction
Successful cross-control of material properties has been reported or proposed for various external stimuli, such as electric field (or electric current),\cite{matsukura15a,parkin08a,yamaguchi04a} magnetic field,\cite{tokura06a,tokura14a} light,\cite{kirilyuk10a} and heat.\cite{koshibae14a} Among them, heat is a particularly important stimulus since it allows for the manipulation of the state of matter through its ability to alter the free energy landscape of a system and to realize a metastable state, depending on cooling kinetics. One such example is a switching phenomenon between (atomic) amorphous and crystal phases as observed in GeTe-related materials,\cite{chen86a,lencer08a,xqliu11a} or between charge-glass and charge-crystal phases in organic materials,\cite{oike15a} where optical reflectivity or electrical resistivity of the system changes significantly, depending on which phase is realized (namely, depending on the cooling speed after the heat injection). This phenomenon in GeTe-related materials is applied as phase-change memory in digital versatile disks (DVD).\cite{jongenelis96a,lencer08a,kolobov14a} 

In addition to the phase-change memory function, GeTe exhibits various intriguing properties, such as a many-valley band structure,\cite{herman68a,ciucivara06a} superconductivity,\cite{cohen64a,hein64a} and thermoelectricity.\cite{snyder08a,levin13a,davidow13a,jklee14a,dwu14a} The $p$-type charge carriers in this system are unintentionally self-doped due to Ge vacancies.\cite{edwards06a} It also exhibits a ferroelectric transition at approximately 700~K, where the structure changes from its high-temperature cubic ($Fm\bar{3}m$; $\beta-$GeTe) to the low-temperature rhombohedral phase ($R3m$; $\alpha-$GeTe),\cite{goldak66a,pawley66a,chattopadhyay87a,rabe87b,schlieper99a,fons10a} with a polar distortion along the cubic [111] direction [see Fig.~\ref{fig1}\,(a)]. Recently the system was predicted\cite{disante13a,picozzi14a}and found\cite{rinaldi14a,krempasky15a} to exhibit a giant Rashba spin-splitting in the bulk, associated with broken inversion symmetry and large spin-orbit interaction. 

GeTe also offers the possibility of enhanced magnetic interactions and applicability in spintronics devices: Magnetism is induced in GeTe when doping with Cr, Mn, or Fe at the Ge site,\cite{rodot66a,cochrane74a,fukuma01a,fukuma03b,knoff11a,tong11a} forming a family of diluted magnetic semiconductors similar with (GaMn)As or (Ga,Mn)N.\cite{jungwirth05a,dietl10a,lchen11a,dietl14a,mtanaka14a} Recently, the possibility of multiferroicity has been claimed in this multifunctional system due to the coexistence of magnetism and ferroelectric distortion.\cite{przybylinska14a} Among the GeTe-based doped systems, \GMT\ has been intensively studied. Single-phase GeTe\,--\,MnTe solid solutions exist over a broad range of concentration up to $x\gtrsim 0.5$.\cite{lechner10a} The partial substitution of Ge$^{2+}$ with isovalent Mn$^{2+}$ reduces the ferroelectric distortion and stabilises the cubic phase. The opposite end member MnTe is an antiferromagnet crystallizing in a different structure with the hexagonal space group $P63/mmc$. The onset temperature \Tc\ of magnetic order was reported to increase linearly with $x$ with maximum values around 165~K for $x = 0.5$.\cite{cochrane74a} More recent works have focussed on thin films of \GMT, finding a carrier-induced enhancement of the \Tc\ values up to 200~K for $x = 0.08$ and hole concentrations of about $1.6\times 10^{21}$~cm$^{-3}$.\cite{chen08a,fukuma08a,lechner10a,hassan11a} The emergence of ferromagnetism in bulk and thin films of \GMT\ has been considered in an RKKY framework plus possible antiferromagnetic correlations by Mn\,--\,Mn direct exchange.\cite{cochrane74a,dietl10a,dietl14a} 

In this Article, we focus on the low-doped region of the phase diagram $x\leq 0.2$ for bulk \GMT, where we found a feature which had been overlooked so far. As shown in Fig.~\ref{fig1}\,(b), there are two distinct magnetic phases, depending on the heat-treatment of the samples, with rather different values of \Tc, called herein low-\Tc\ and high-\Tc\ phase. The origin of the high-\Tc\ phase was identified as the formation of Mn-rich / Mn-poor regions in terms of state-of-the-art synchrotron x-ray diffraction and energy dispersive x-ray analysis. We also successfully demonstrated that samples can be repeatedly switched from the low-\Tc\ into the high-\Tc\ phase and vice versa by changing the heat treatment. The latter adds another interesting feature to the multifunctional semiconductor GeTe, namely magnetic phase-change-memory functionality.

%Results
\section*{Results}
\begin{figure}[t]
\centering
\includegraphics[width=14cm,clip]{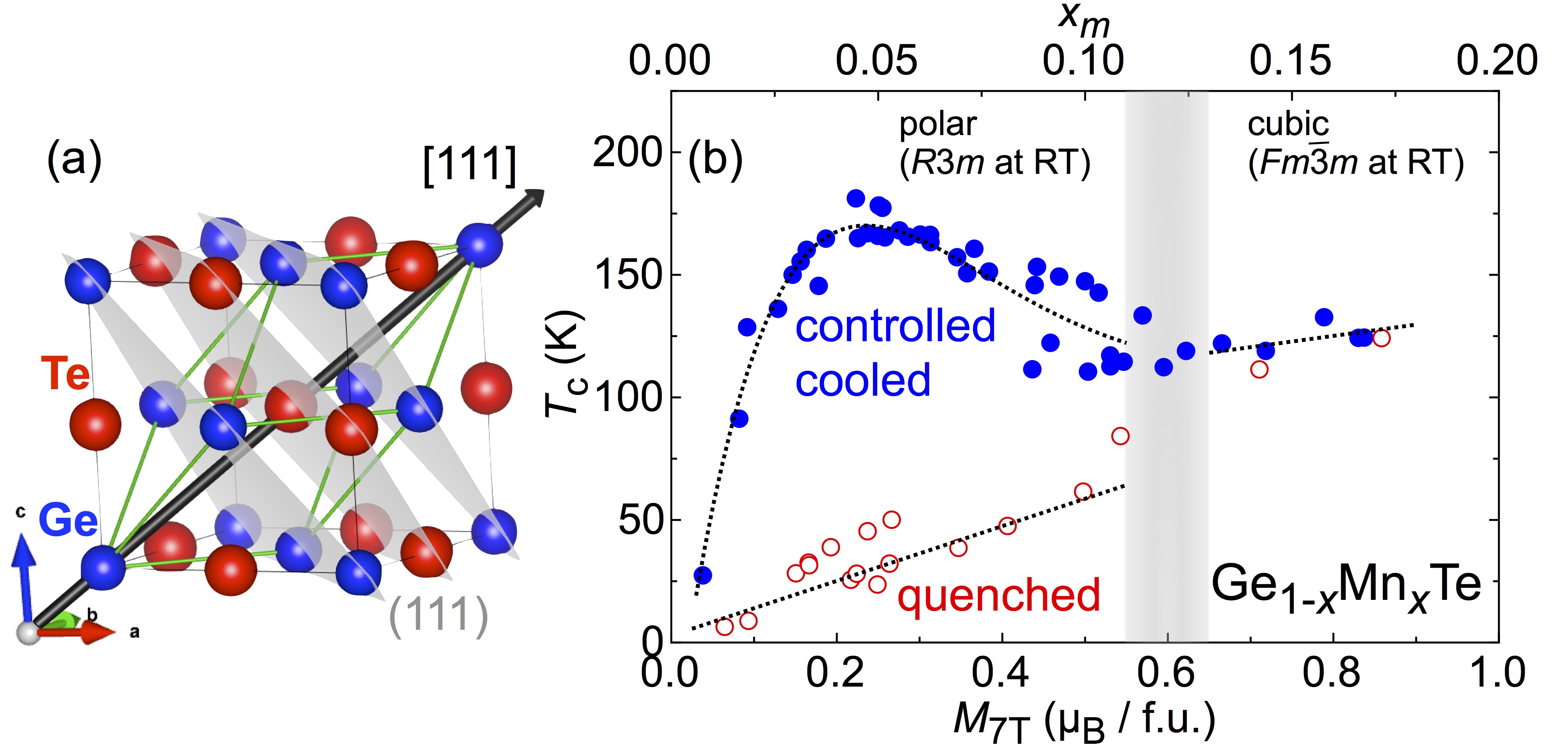}
\caption{Structure and phase diagram of \GMT: (a) Structural plots depicting the high-temperature cubic and the low-temperature rhombohedrally-distorted GeTe structure superimposed in its pseudo-cubic setting. The high-temperature structure is of rock-salt type. The atoms Ge (blue), Te (red), and the crystallographic (111) planes (grey) are indicated. The black arrow depicts the cubic [111] direction, along which the polar distortion occurs. (b) Magnetic phase diagram of \GMT. Horizontal axis is magnetisation value at 2~K and 7~T, which is an effective measure of the Mn concentration. The corresponding number $x_m$ is plotted at the upper horizontal axis. Filled symbols (blue) refer to the onset of ferromagnetism in controlled-cooled samples, open symbols (red) to quenched samples; see the Methods section for the details of the heat-treatment procedures. The grey shaded area highlights the $x_m$ range where the structural phase transition from rhombohedral to cubic phase drops below room temperature. Dotted lines are guides to the eyes.}
\label{fig1}
\end{figure}

The magnetic phase diagram of \GMT\ based on our present results is shown in Fig.~\ref{fig1}\,(b). The ferromagnetic transition temperature \Tc\ is plotted against the magnetic moment at $B=7$~T and $T=2$~K as extracted from field-dependent magnetisation measurements (see below). The corresponding number $x_m$ are shown on the upper horizontal axis as an effective measure of the Mn concentration. Here, $x_m$ is calculated under the assumption that each Mn$^{2+}$ ion with $S=5/2$ contributes with its full moment 5~$\mu_{\rm B}$. The magnetic phase diagram can be divided into two sections (i) and (ii): (i) Below $x_m\sim 0.12$, there are two distinct magnetic phases with different values of \Tc. Which magnetic phase is realised depends on the heat treatment of a \GMT\ batch during growth. Quenching from the cubic phase (at 900~K) into water leads to the formation of the low-\Tc\ phase [red open symbols in Fig.~\ref{fig1}\,(b)] while a slow and controlled cool down to room temperature, typically 5~K/h or less, establishes the high-\Tc\ phase (blue filled symbols). In the low-\Tc\ phase, the onset of ferromagnetism increases linearly with $x$. In contrast, the high-\Tc\ phase exhibits a dome-like shape with values of \Tc\ differing by a factor as large as five to six, compared to the low-\Tc\ phase around $x_m\sim 0.05$, i.e., where the maximum $\Tc\approx 180$\,K is achieved. (ii) Above $x_m\sim 0.12$, the two different phase boundaries merge, and upon further increasing $x_m$, we do not observe apparent differences any more between samples from batches heat-treated in either way. As in the low-\Tc\ phase, \Tc\ increases linearly with $x_m$, although the slope is somewhat smaller than observed for $x_m\leq 0.12$. 
%Figure 1

\begin{figure}[t]
\centering
\includegraphics[width=12cm,clip]{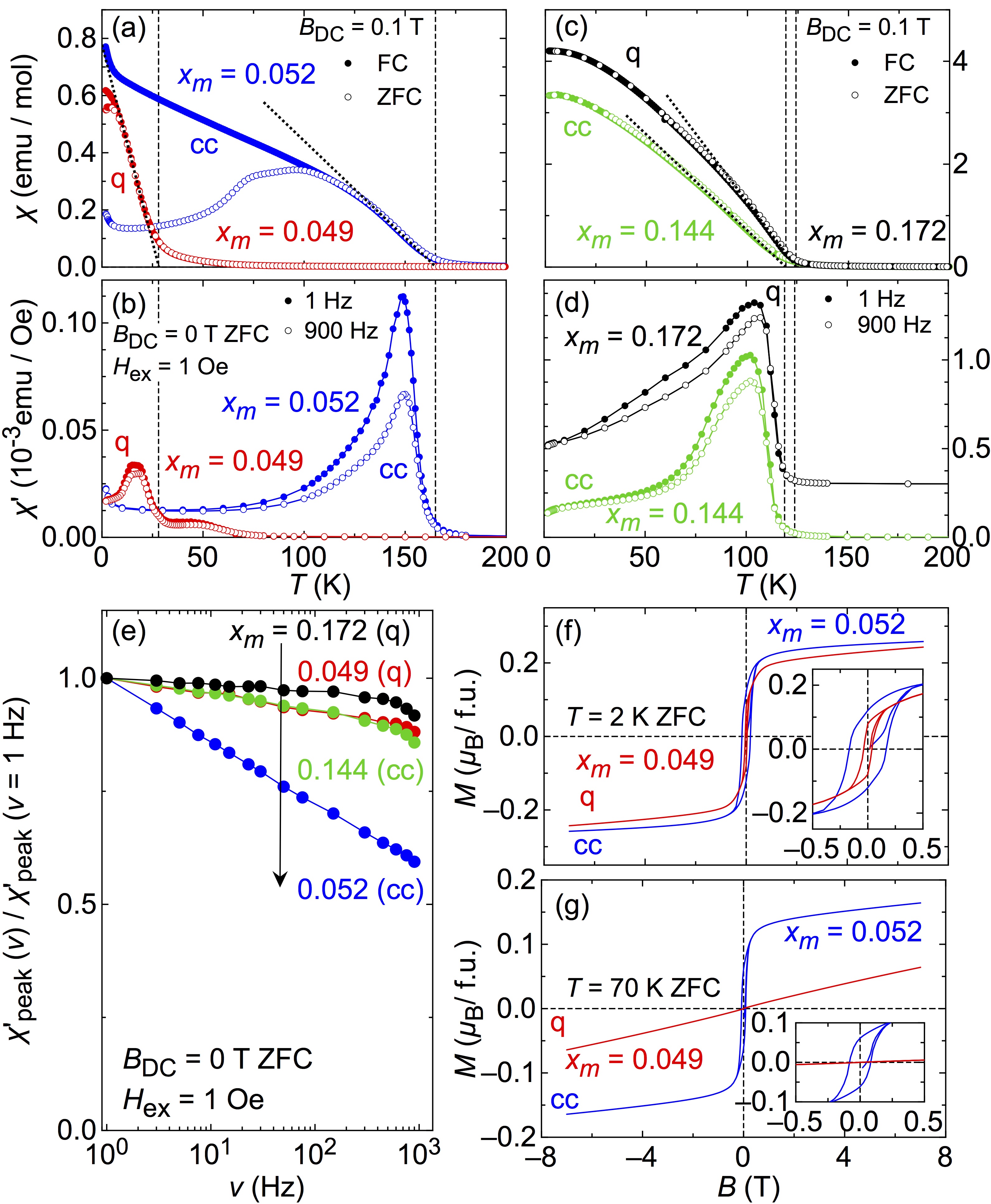}
\caption{Magnetisation data of \GMT: (a) DC- and (b) AC-susceptibility data for a controlled-cooled ($x_m = 0.052$; blue symbols) and a quenched sample ($x_m=0.049$, red symbols) in the low-$x_m$ region of the phase diagram. (c) and (d) show the respective data for samples with higher $x_m$ (controlled cooled: $x_m=0.144$, green symbols, and quenched: $x_m=0.172$, black symbols). In panels (a) and (c) filled symbols are measured in field-cooling (FC) runs and open symbols denote data measured after zero-field cooling (ZFC). The dotted lines are fits to the data and show how the magnetic transition temperatures \Tc\ were estimated. The vertical lines in (a)\,--\,(d) indicate thus determined \Tc\ values. The AC-susceptibility data in panels (b) and (d) were taken in zero DC magnetic fields and an AC excitation field of 1~Oe. Closed symbols in both panels refer to data taken at an excitation frequency $\nu$ of 1 Hz (lowest applied $\nu$), open symbols to data taken in 900 Hz (largest applied $\nu$). The data of the quenched sample in (d) are shifted vertically for clarity. In (e), the frequency dependence of the normalized peak value $\chi$' for all four samples are summarized, with the horizontal axis in logarithmic scale. Panels (f) and (g) show field-dependent magnetisation data at 2~K and 70~K, respectively. The insets are an expanded view around the origin.}
\label{fig2}
\end{figure}

Figure~\ref{fig2} summarises and compares DC- and AC-susceptibility data of samples in the low-$x_m$ region in panels (a) and (b), for a controlled-cooled (labelled `cc') and a quenched (`q') sample. In the DC-susceptibility data of the lower-$x_m$ samples in (a), the difference in the onset temperature \Tc\ of ferromagnetism (as indicated by vertical dashed lines) between the controlled-cooled ($x_m = 0.052$, $\Tc=165$\,K) and the quenched sample ($x_m=0.049$, $\Tc=28$\,K) is as much as six times despite the very similar values of $x_m$, which is clearly exemplified by an almost identical saturation moment of the ferromagnetic hysteresis as shown in panel (f) at $T = 2$\,K. The difference in \Tc\ is reflected in the very different behaviour of the hysteresis curves at $T = 70$\,K [panel (g)], since at this temperature the quenched sample is already in its paramagnetic state. Another eminent difference in the temperature-dependent DC-susceptibility of the low-$x_m$ samples is that the zero-field cooled (ZFC) and field-cooled (FC) data for the quenched sample are identical except at very low temperatures while there is a large difference in the case of the controlled-cooled sample below roughly $2\Tc/3$. The latter resembles the behavior of a spin-glass or cluster-glass like magnetic phase while the former is closer to conventional ferromagnetic order. 

To further inspect this observation, AC-susceptibility data measured on the same samples are shown in panel (b). It was measured in zero external field and at an AC excitation field of 1\,Oe for various excitation frequencies 1\,Hz\,$\leq \nu \leq 900$\,Hz. The data obtained at the lowest and the largest frequency are shown (filled symbols: $\nu = 1$\,Hz, open symbols: $\nu = 900$\,Hz) for the purpose of clarity. Just below \Tc, a clear maximum is observed for both the controlled-cooled and the quenched sample. The frequency dependence is much more pronounced in  controled-cooled than quenched samples.

Panels (c) and (d) contain the equivalent data measured under the same conditions for a controlled-cooled ($x_m=0.144$, $\Tc=119$\,K) and a quenched sample ($x_m=0.172$, $\Tc=124$\,K) from the large-$x_m$ part of the phase diagram. Vertical dashed lines in both panels indicate only slightly differing \Tc\ values of each sample, which is in sharp contrast to the case of low-$x_m$ samples. Here the FC and ZFC magnetisation data of both samples do not show any significant difference down to the lowest measurement temperature. We note that the field-dependent data taken on both higher-doped samples exhibit qualitatively similar $M(B)$ curves (not shown) as the quenched low-$x_m$ sample. Namely, all larger-$x_m$ samples exhibit smaller hysteresis loops similar to the low-$x_m$ quenched samples while controlled-cooled low-$x_m$ samples show the largest hysteresis. The AC-susceptibility data in (d) exhibits peaks below \Tc, the heights of which are frequency dependent for both samples.

For a better comparison, the normalized values of the peaks are replotted against the frequency in panel (e) on a logarithmic scale. Apparently the frequency dependence is much stronger for the controlled-cooled low-$x_m$ sample which shows a peak suppression of about 40\,\%. In contrast, the three other samples exhibit a very similar frequency dependence and the respective suppression of the peaks is less than 15\,\%. A frequency dependence of the peak heights in AC susceptibility is a characteristic feature expected for a glass-like magnetic state.\cite{binder86a} Another common approach to analyse such AC susceptibility data is to examine the frequency dependence of the peak temperature $T_{\rm max}$.\cite{tholence80a,aruga88a} We estimated $T_{\rm max}$ for the controlled-cooled sample with $x_m=0.052$ shown in Fig.~\ref{fig2}~(b) (blue data symbols) for all measured frequencies $1~{\rm Hz} \leq \nu \leq 900$~Hz and fitted the Fulcher law $\nu=\nu_0 \exp[-E_{\rm a}/k_{\rm B}(T_{\rm max}(\nu)-T_0)]$ to these data as described in Ref.~\onlinecite{tholence80a}. Although the fit result (not shown) is not perfect at low frequencies, it describes well the data at higher frequencies and resembles the behavior reported for metallic RKKY-spin glasses rather than systems in which the spin glass phase emerges due to geometrical frustration.\cite{tholence80a}

All the AC- and DC-magnetic data taken together, the controlled-cooled low-$x_m$ sample exhibits a more spin-glass-like or clustered magnetic structure while the magnetic phases of both larger-$x_m$ and the low-$x_m$ quenched samples are similar to an ordinary ferromagnet. One possible scenario is a segregation of the magnetic Mn ions in the low-$x_m$ high-\Tc\ sample during the controlled-cooling process. The high-\Tc\ phase is characterized by Mn-rich islands or clusters embedded into a lake or matrix of relatively Mn-poor GeTe, as it is expected if a spinodal decomposition occurs,\cite{ksato05a,dietl10a,dietl15a}while the low-\Tc\ phase consists of a more homogeneous Mn distribution.
%Figure 2

\begin{figure}[t]
\centering
\includegraphics[width=12cm,clip]{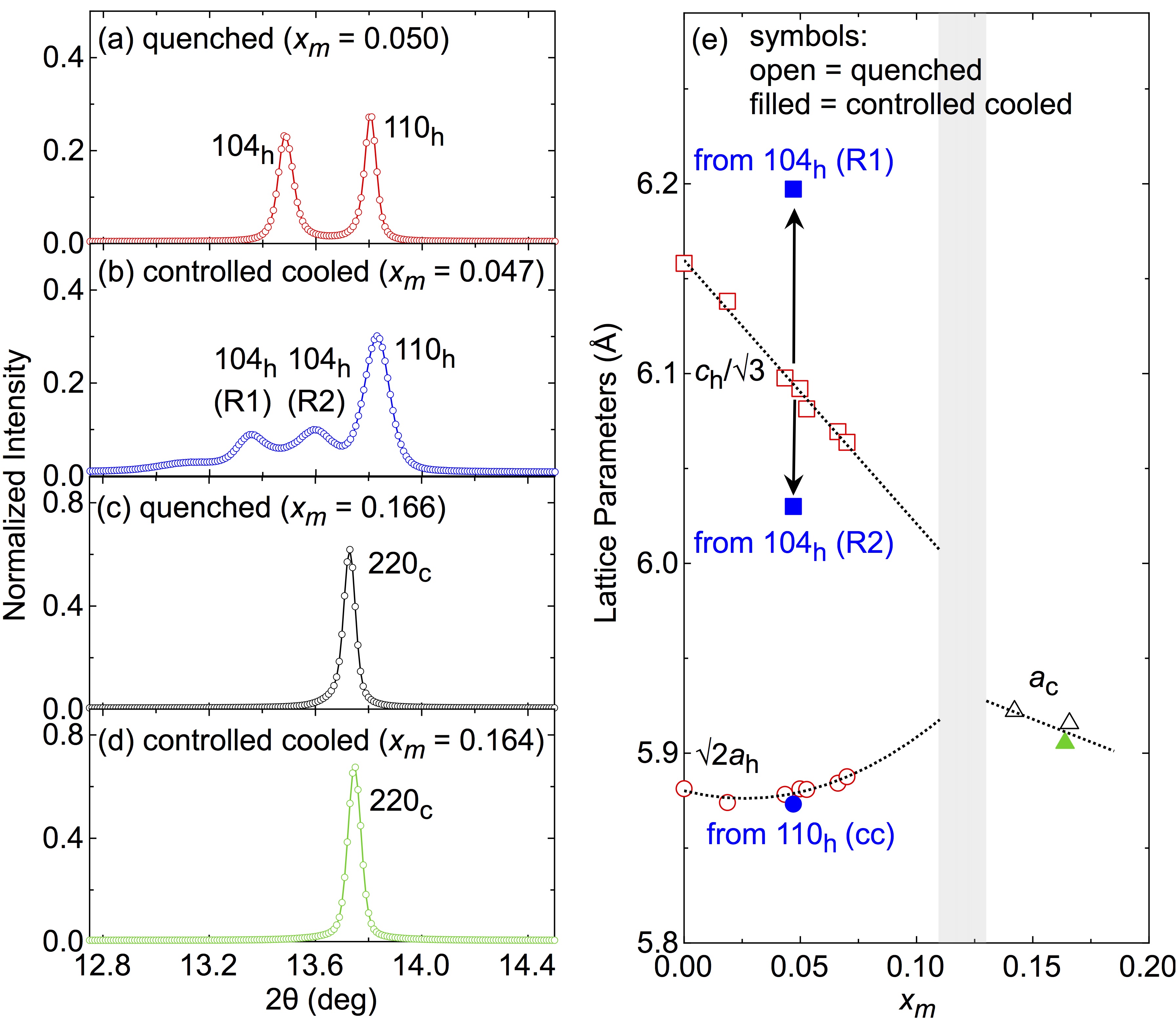}
\caption{XRD data of \GMT: Data are shown for four different samples around the single cubic (220) peak and the corresponding (104) and (110) reflections (hexagonal setting) that are split due to the polar distortion: (a) quenched sample with $x_m = 0.050$, (b) controlled-cooled sample with $x_m = 0.047$, (c) quenched, $x_m = 0.166$, and (d) controlled cooled, $x_m=0.164$. (e) Lattice constants as estimated from the XRD data for these and additional quenched samples. Filled symbols refer to controlled-cooled, open symbols to quenched samples. The two data points labelled $R1$ and $R2$ refer to the $c_{\rm h}$ lattice constants estimated for the two $104_{\rm h}$ peaks labeled in the same way in panel (b), see text. The filled circle denotes the corresponding $a_{\rm h}$ lattice constant. The lattice parameters in hexagonal setting $a_{\rm h}$ and $c_{\rm h}$ transform into the pseudo-cubic setting ($\tilde{a}_c$, $\tilde{c}_c$) as follows: $\tilde{a}_c =\sqrt{2}a_{\rm h}$ and $\tilde{c}_c = c_{\rm h}/\sqrt{3}$, and $\sqrt{2}a_{\rm h}$ and $c_{\rm h}/\sqrt{3}$ are plotted for a better comparability.} 
\label{fig3}
\end{figure}

To test this scenario and further characterize the different magnetic phases, we carried out a high-resolution x-ray diffraction study employing synchrotron radiation with the wavelength of 0.5001(1)~\AA. The main result is shown in Fig.~\ref{fig3}. As before, controlled-cooled and quenched powder samples from both the low-$x_m$ and the large-$x_m$ sections in the phase diagram were analysed: The differences in the magnetic phases are reflected in differences in the XRD patterns. Figures~\ref{fig3}\,(a) to (d) show XRD data for four selected samples on a magnified view of the cubic $220_{\rm c}$ reflection around $2\theta\sim 13.5\,^{\circ}$: (a) $x_m= 0.050$, $\Tc=24$\,K, quenched; (b) $x_m= 0.047$, $\Tc=171$\,K, controlled cooled; (c) $x_m= 0.166$, $\Tc=125$\,K, quenched; (d) $x_m= 0.164$, $\Tc=127$\,K, controlled cooled.  In quenched low-$x_m$ samples as shown in Fig.~\ref{fig3}\,(a), we observe two comparably sharp peaks indexable as the $104_{\rm h}$ and $110_{\rm h}$ reflections in hexagonal setting, which are expected for the polar rhombohedrally-distorted GeTe phase. For large-$x_m$ samples [Fig.~\ref{fig3}\,(c) and (d)] above the structural phase transition [shaded areas in Figs.~\ref{fig1}\,(b) and \ref{fig3}\,(e)], there is only one comparably sharp peak, irrespective of the heat treatment, which is indexable as the corresponding $220_{\rm c}$ reflection in the cubic GeTe phase. This is in agreement with the observation that for $x_m > 0.12$, there is no clear difference in the magnetic state any more between controlled-cooled and quenched samples. 

However, for controlled-cooled samples with smaller $x_m$, i.e., below the grey-shaded area in the phase diagram, the situation turns out to be much more complicated. As can be seen in panel (b), the $104_{\rm h}$ reflection has split into two broader peaks with lower intensity.  We note that we also observe similar splittings of other reflections $(hkl_h)$ with non-zero $l_h$ value, indicating that this splitting is not due to an impurity phase. At the same time the $110_{\rm h}$ peak also broadens. Keeping in mind that the $104_{\rm h}$ reflection provides information about the degree of the rhombohedral distortion, we interpret the apparent double-peak structure as an indication that controlled-cooled low-$x_m$ sample consists of  domains with different degrees of rhombohedral distortion while obeying the same overall crystal symmetry. In the present case, we assume domains with two main distortions and label the two peaks in Fig.~\ref{fig3}\,(b) as `$R1$' and `$R2$' for simplicity. The former domain $R1$ is strongly distorted, close to the situation in pristine GeTe while the latter exhibits a smaller distortion which implies that the $R2$ component has a larger Mn concentration, naturally supporting the aforementioned scenario of a slow-cooling triggered spatial inhomogeneity of the Mn distribution. 

In Fig.~\ref{fig3}\,(e), the estimated lattice constants for these four samples and additional quenched samples are plotted against $x_m$ in pseudo-cubic setting for a better comparability. The $c_{\rm h}$ parameter of quenched samples shrinks pronouncedly while $a_{\rm h}$ increases slightly with $x_m$. In the `real' cubic phase for $x_m\gtrsim0.12$, the lattice parameter shrinks. In the controlled-cooled low-$x_m$ sample [Fig.~\ref{fig3}\,(b)], the single lattice parameter $a_{\rm h}$ (common to the $R1$ and $R2$ phases) as estimated from the $110_{\rm h}$ reflection is plotted with a filled ball symbol, fitting into the systematic change of the $d$ spacing perpendicular to the polar axis. Due to the peak splitting, we estimated the lattice parameters $c_{\rm h}$ for each of the two $104_{\rm h}$ reflections $R1$ and $R2$ as denoted by filled square symbols in panel (e). The difference in the degree of the rhombohedral distortion is reflected in two very different lengths of $c_{\rm h}$. As expected, above the structural phase transition there is no difference between samples treated by either cooling recipe. 

%Figure 3

\begin{figure}[t]
\centering
\includegraphics[width=12cm,clip]{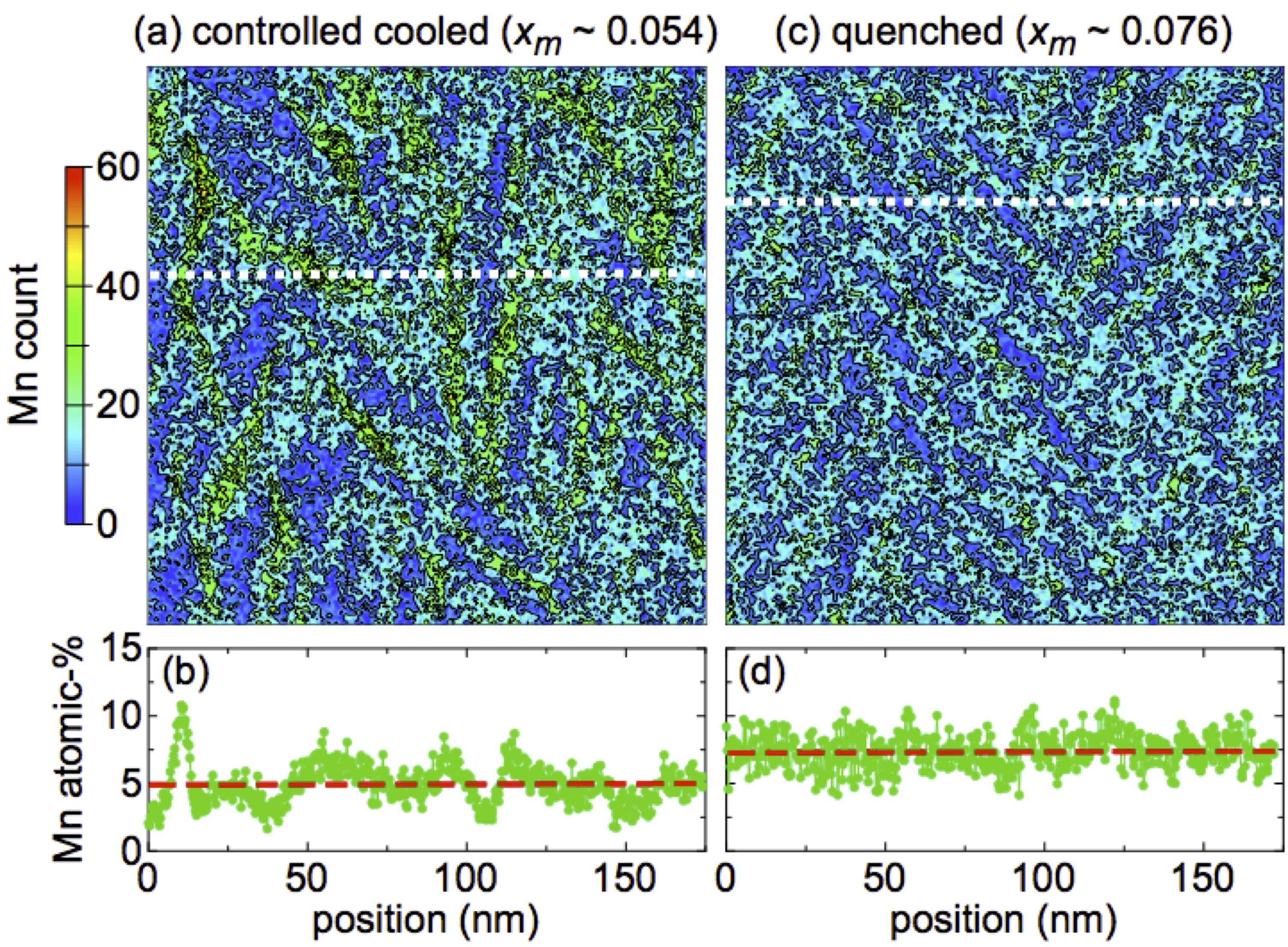}
\caption{EDX data of \GMT: Comparison of EDX images (Mn count) of (a) a controlled-cooled sample with a Mn concentration of $x_m = 0.054$ and (b) a quenched sample, $x_m=0.076$. The size of the view area of (a) and (c) is $170\times 170$~nm$^2$. The white dashed lines denote the approximate position of the line scans shown in panels (b) and (d), respectively, for the two samples. In (b) and (d), the average Mn concentrations along these line scans are indicated by a red dashed line, which show good agreement with the respective $x_m$ values. The controlled-cooled sample exhibits a more inhomogeneous Mn distribution than the quenched sample, see text.} 
\label{fig4}
\end{figure}

As a next step, we carried out an energy-dispersive x-ray (EDX) analysis by using a scanning transmission electron microscope (STEM) to probe and visualize the spatial Mn distribution in two samples from the low-$x_m$ section of the phase diagram. The resulting EDX mappings of Mn count are shown in Fig.~\ref{fig4}\,(a) for a controlled-cooled sample ($x_m= 0.054$, $\Tc=167$\,K) and in (c) for a quenched sample ($x_m= 0.076$, $\Tc=45$\,K). Figures~\ref{fig4}\,(b) and (d) show the average atomic percentage of Mn ions from ten line scans taken around the white dashed lines in panels (a) and (c). Clearly, the controlled-cooled sample exhibits a much more inhomogeneous Mn distribution than the quenched sample, although the quenched sample also shows a slight inhomogeneity. This finding is in accord with the observations made in relation to magnetic and XRD data. The characteristic length scale of the Mn clustering amounts to several tens of nm, as seen in Figs.~\ref{fig4}\,(a) and (b).

%Figure 4

\begin{figure}[t]
\centering
\includegraphics[width=10cm,clip]{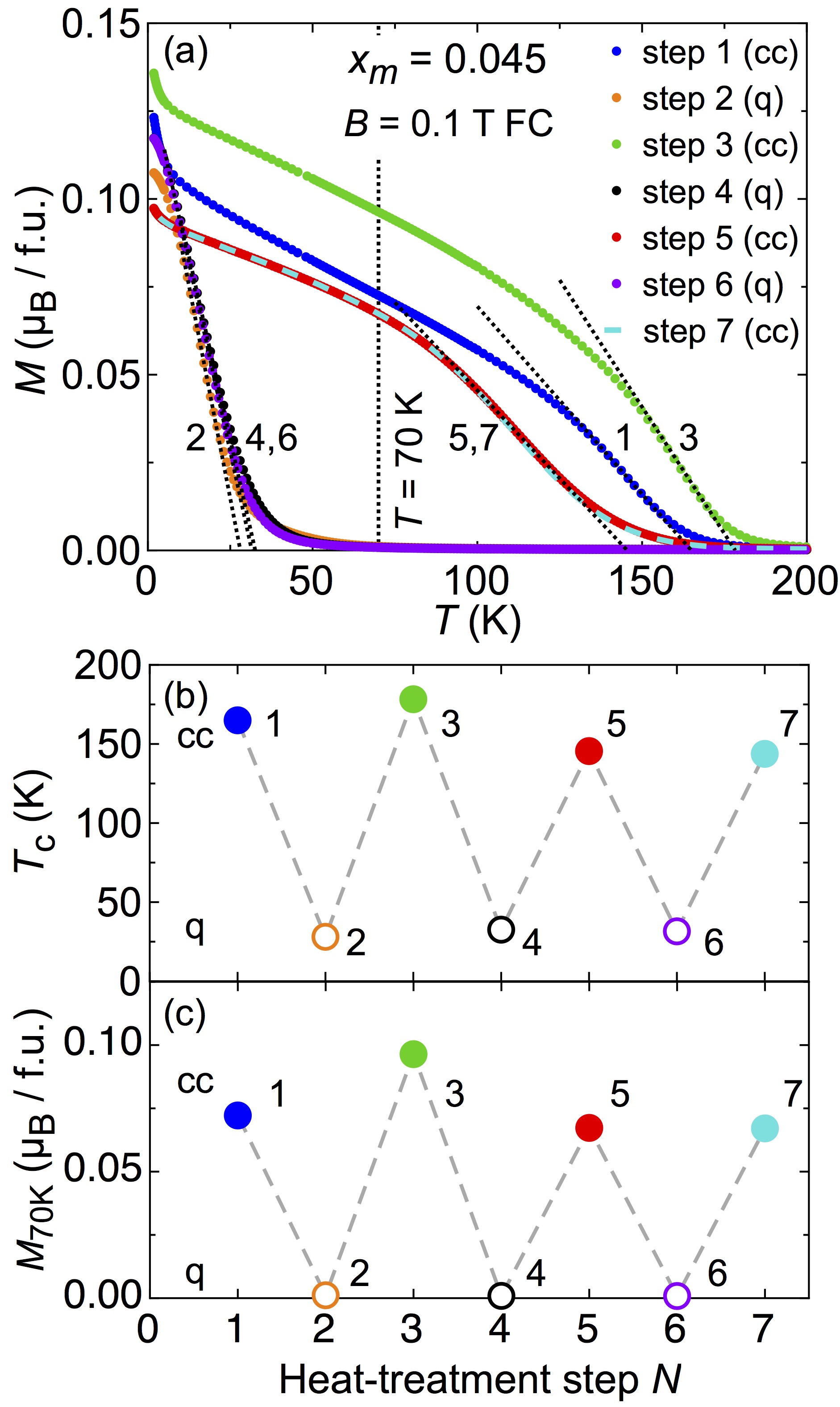}
\caption{Demonstration of phase switching in \GMT: (a) summarizes DC magnetisation data for each of the seven phase-conversion steps (controlled cooled `cc' vs.\ quenched `q'). All data are taken on the same specimen upon field cooling in $B=0.1$~T. The Mn concentration as estimated from the magnetisation value at 2~K and 7~T is $x_m=0.045$. The dotted lines are fits to the data and show how the magnetic transition temperatures \Tc\ were determined. The numbers 1 to 7 indicate the chronological order of the heat treatment and hence measurements. The variation of \Tc\ and the magnetisation at 70~K as indicated by the vertical line in (a) among the different switching steps are shown in (b) and (c), respectively.}
\label{fig5}
\end{figure}

Another important issue is whether the different magnetic phases with high-\Tc\ and low-\Tc\ values in the low-$x_m$ section of the magnetic phase diagram can be repeatedly switched back and forth. To demonstrate this feature, a sample from an initially controlled-cooled batch ($x_m=0.045$) was chosen and six times (from step 2 to step 7) switched as shown in Fig.~\ref{fig5}\,(a). The  \Tc\ values are replotted in panel (b) as a function of the heat-treatment step, along with the magnetisation value at 70~K in Fig.~\ref{fig5}\,(c) as indicated by the vertical line in panel (a). For either data set, open symbols refer to measurements on quenched and filled symbols to controlled-cooled samples. This finding indicates that both observed magnetic phases are reproducibly switchable into each other, fulfilling one of the essential requirements for phase-change-memory functionality. It should be noted that the slight variation in \Tc\ of the high-\Tc\ phase between the different phase switching steps 1, 3, 5, and 7 shown in Fig.~\ref{fig5} is probably a consequence of slightly different degrees of Mn inhomogeneity in the sample obtained after each thermal cycle, rather than due to a degradation in the bulk of the sample.
%Figure 5

% Discussion
\section*{Discussion} The present data provide a consistent picture suggesting that the difference in the strength of ferromagnetic interaction originates from the degree of clustering or inhomogeneity of the substitutionally doped Mn$^{2+}$ ions. On the one hand, high \Tc\ values are found when the Mn-rich clusters or islands in a lake of Mn-poor, almost pristine GeTe have the time to arrange themselves during the controlled-cooling process, i.e., for $x_m$ around 0.05 leading to a highly inhomogeneous situation. Within the Mn-rich region, the ferromagnetic interaction is stronger than expected for the averaged Mn concentration within the RKKY scheme. On the other hand, when the Mn ions are not able to cluster, i.e., when the system is quenched from the high-temperature homogeneous arrangement, the \Tc\ values turn out to be much smaller. Apparently, the homogeneous situation is frozen and the magnetic interaction is weaker. Upon further doping, the Mn concentration becomes large enough to allow for cubic phase fractions to emerge. This leads to a suppression of \Tc\ when approaching the critical $x_m$ range around $\approx 0.12$ where the structural phase transition drops below room temperature. Hence the high-\Tc\ and low-\Tc\ phase boundaries merge and the clustering is reduced. One might expect that at sufficiently large Mn-doping concentrations, similarly large \Tc\ values must be achievable. In fact, the maximum \Tc\ was reported in the old literature to be around 160~K for $x \sim 0.5$,\cite{cochrane74a} somewhat smaller than $\Tc=180$~K reported in this Article. Assuming that the slope of the phase boundary above $x_m\sim 0.12$ holds and extrapolating the phase line towards larger $x_m$ values, a comparably large \Tc\ of 180~K is expected for $x_m\sim 0.4$ in the cubic phase. However, even the local spatial fluctuation of the Mn content for the low-$x_m$ and high-\Tc\ sample [see Fig.~\ref{fig4}\,(b)] seems not to reach such a high value as $x_m\sim 0.4$ (cubic). This discrepancy may have the important implication that the possible accommodation of relatively high-$x_m$ ($\gtrsim 0.1$) Mn content in the rhombohedrally distorted polar lattice structure (e.g., the $R2$ phase in Fig.~\ref{fig3}) may host higher \Tc\ values than the comparably or even higher Mn-doped cubic phase. The characteristic Rashba-type spin-split valence band structure in the polar state, as recently proven theoretically\cite{disante13a,picozzi14a} and experimentally,\cite{rinaldi14a,krempasky15a} may play some role in giving rise to such a large difference in RKKY interactions in the polar and cubic lattices. 

The appearance of Mn inhomogeneity and different rhombohedral distortions in this system suggest that a spinodal decomposition occurs in the low-Mn-concentration region when a sample is cooled down slowly: A uniform solid solution becomes unstable against composition modulations upon cooling.\cite{ksato05a,dietl15a,hiroi13a} We note here that this process is totally different from the structural phase change between crystalline and amorphous. The highest temperature which the samples experience during the heat-treatment process is 900~K, which is well below the melting temperature of \GMT. Such a nanoscale phase separation is discussed in related compounds, and means that the spin subsystems undergo a segregation, i.e., regions with an either high or low concentration of the magnetic dopant are formed. It is also known that the degree of inhomogeneity in diluted magnetic semiconductors can indeed influence the strength of the ferromagnetic interaction and strong inhomogeneity may significantly increase \Tc.\cite{ksato05a,jamet06a} 
Moreover it was reported that the ferromagnetic transition temperature can vary depending on the heat-treatment, as, e.g., in the text-book diluted magnetic semiconductor (Ga,Mn)As,\cite{jungwirth05a} but the mechanism is different from the spinodal decomposition proposed here for. In the case of (Ga,Mn)As, the annealing largely affects the amount of interstitial Mn ions, resulting in the difference of $T_{\rm c}$. In the present case, however, the almost identical properties in the high-$x_m$ region among the samples treated in either thermal process indicate that interstitial Mn ions, if any, play a rather minor role in governing the magnetic properties (although the effect of interstitial Mn ions should be investigated in detail in the future). Therefore, the magnetic phase change functionality reported here was never identified before.

It was theoretically found that nanoscale spinodal decomposition in diluted magnetic semiconductors can lead to large values of \Tc\ in cases where there is only a short-range magnetic exchange interaction.\cite{ksato05a,katayamayoshida07a} We speculate that the high-\Tc\ values reported for thin films\cite{chen08a,fukuma08a,hassan11a} are actually belonging to the here-reported bulk high-\Tc\ phase (while the low-\Tc\ phase line is probably the one which was observed and reported in the old literature Ref.~\onlinecite{cochrane74a}). High \Tc\ values are only reported for thin films with small $x$ which were grown at comparably low temperatures $<620$~K,\cite{fukuma08a,hassan11a} i.e., in the same temperature window which was found in this study to trigger the formation of the high-\Tc\ phase in our samples. Therefore the proposed spinodal decomposition mechanism is probably the intrinsic origin for the existence of the high-\Tc\ magnetic phase in \GMT. Another interesting issue is the size of the Mn clusters formed due to the spinodal decomposition. For bulk \GMT\ we estimated an average Mn cluster size to be several tens of nanometers by employing Scherrer`s formula to the broadened XRD patterns of controlled-cooled powder samples with large \Tc\ values. We also estimated the mean-free path of a sample with $x \approx 0.09$, high-\Tc\ phase, from preliminary transport measurements. The mean-free path is smaller than 10~nm and this makes sense since the enhancement of \Tc\ could not occur if the Mn cluster size were smaller than the mean-free path and the inhomogeneity of the Mn concentration were averaged out. This is an interesting starting point for future studies on the ferromagnetism realised in \GMT.

One might speculate whether it could also be possible to gain control over the magnetic-phase-switching process by electronic means, i.e., whether it is possible to switch the magnetic phases by electric fields utilising the ferroelectric distortion which sets in upon cooling through $\sim 700$~K. Unfortunately the semiconductor GeTe is a fairly good metal with room-temperature values of the longitudinal resistivity of a few 100~$\muup\Omega$cm and unintentionally self-doped charge carrier concentrations $n$ of the order of $10^{21}$~cm$^{-3}$. To drive the magnetic phase change by electrical fields, one has to reduce $n$. Mn doping alone does not seem to reduce $n$ sufficiently, at least in the doping range in question, i.e., $x_m \lesssim 0.12$. In general, the role of the density of the self-doped charge carriers in \GMT\ remains an open question. In diluted magnetic semiconductors, the ferromagnetic interaction is generally believed to depend on the charge carrier concentration,\cite{katayamayoshida07a, dietl14a} which was also reported for thin films of \GMT.\cite{fukuma02a}

% Summary
%\section{Summary}
In conclusion we report the finding that a different heat treatment (quenching vs.\ slow cooling) results in either of two competing ferromagnetic phases with significantly different ordering temperatures \Tc\ in the bulk diluted magnetic semiconductor \GMT. The high-\Tc\ phase partly resembles a spin- or cluster-glass state with maximum ordering temperatures of up to $\sim 180$\,K around $x_m\sim 0.05$ which had not been reported before. It is characterized by an inhomogeneous distribution of the doped Mn$^{2+}$ ions due to a spinodal decomposition taking place upon slow cooling from above the ferroelectric transition temperature ($\sim700$~K). The low-\Tc\ phase  is more homogeneous and closer to conventional ferromagnetic order. The two phases merge around $x_m \approx 0.12$. At the same time the ferroelectric lattice distortion vanishes. Moreover, it was demonstrated that repeated switching back and forth between the two distinct phases is possible by either quenching (high-\Tc\ to low-\Tc) or controlled cooling (low-\Tc\ to high-\Tc). This adds another interesting feature to the intriguing semiconductor GeTe in terms of a magnetic phase-change-memory functionality. 

%Experiment: Samples and Methods
\section*{Methods}
\textbf{Sample Preparation and Characterization.} 
Polycrystals of \GMT\ for nominally $0\leq x< 0.2$ were grown by conventional melt growth and Bridgman methods. Stoichiometric mixtures of GeTe (purity: 5N) and MnTe (3N+) were thoroughly mixed and sealed into evacuated quartz glass tubes. In the conventional melt growth runs, the batches were heated to 1073 - 1123~K (melting point of GeTe: $T_{\rm m} \approx  1000$~K;  upon Mn doping it gradually increases to $T_{\rm m}\approx 1073$~K for $x\approx 0.5$.\cite{johnston61a}), kept there for 12\,--\,24~h and subsequently cooled down to approximately 900~K which is still in the cubic high-temperature phase for all samples examined here (GeTe: $T_{\rm struct}\approx 700$~K; upon Mn doping $T_{\rm struct}$ decreases). Then the batches were (i) slowly cooled (5~K/h) to room temperature or (ii) quenched into water. In the Bridgman growth method, the upper heater was set to 1123~K and the lower to 623~K. The mixed powder was kept at the upper heater's temperature for 12\,--\,24~h. Then the batch was slowly lowered (2~mm/h) from the upper heater towards the lower heater and again (i) slowly cooled down, or (ii) the quartz tubes were quenched when the batch position corresponded to approximately 900~K. We tried different annealing times at 900~K without finding any impact on the magnetic phase, which implies that the relevant temperature range for the clustering process during the slow-cooling process is below that temperature. There is always a slight gradient of the Mn concentration in batches grown by either recipe / method. Therefore, for small samples cut or broken from an as-grown batch, the magnetization moment at $T=2$~K and $B=7$~T is used to estimate the Mn concentration effectively. This approach was verified for selected samples by chemical composition determination using a SEM-EDX apparatus (JEOL JCM-2000).

\textbf{Measurement.}
DC-magnetisation and AC-susceptibility data were measured with commercial magnetometers (MPMS XL and MPMS-3, Quantum Design). The synchrotron radiation experiments were performed at BL44B2 in SPring-8 with the approval of RIKEN (Proposal No.\ 20150045), and with a commercial in-house apparatus (RIGAKU). Scherrer's formula $L=K\lambda/(\Delta_{2\theta}\cos(\theta))$ was used to estimate the Mn cluster size $L$ from the peak width. Here $K\approx 1$ is a form factor, $\lambda$ the wave length of the used radiation, and $\theta$ the Bragg angle. STEM-EDX data was taken at JEOL Ltd.\ by employing a JEM-2800 apparatus. The onset temperature of ferromagnetism is defined as the initial linear slope as indicated in Figs.~\ref{fig2}\,(a) and (c) and \ref{fig5}\,(a) by dashed lines.
The software VESTA was used for the structure plot in Fig.~\ref{fig1}\,(a), see Ref.~\onlinecite{momma11a}.
 
%\section*{References}
%\bibliographystyle{naturemag}
%\bibliographystyle{apsrev4-1}
%\bibliography{/PaperBase/preload,/PaperBase/Chalcogenides,/PaperBase/GeTe,/PaperBase/Thermoelectrics,/PaperBase/SiC-Si-C,/PaperBase/Superconductivity,/PaperBase/sonstigePaper,/PaperBase/Lehrbuch,/PaperBase/TopolIns,/PaperBase/Skyrmions,additionalbib}


% acknowledgement
\section*{Acknowledgments}
This work is supported by a Grants-in-Aid for Scientific Research (S) from the Japan Society for the Promotion of Science (JSPS, No. 24224009). MK is supported by a Grants-in-Aid for Young Scientists (B) (JSPS, KAKENHI No. 25800197) and by a Grants-in-Aid for Scientific Research (C) (JSPS, KAKENHI No. 15K05140).

%\section*{Author contributions}
%M.K. and Y.K. carried out the sample preparation. M.K. performed the magnetisation measurements supported by A.K.; M.K., T.N., and K.K. performed the synchrotron XRD measurements with support from M.T.; X.Z.Y. and N.E. performed the TEM-EDX measurements. M.K. and T.N. analysed the data and M.K. wrote the manuscript. All coauthors read the manuscript and commented. M.K., T.N., X.Z.Y., T.A., Y. Taguchi, and Y. Tokura jointly discussed the results.

%\section*{Additional information}

%\textbf{Competing financial interests: }The authors declare no competing financial interests.

\end{document}